\documentclass[useAMS,usenatbib]{mn2e}

\usepackage{graphicx}
\usepackage{subfigure}

\def\kms      {\ifmmode{\rm km\,s}^{-1} \else km\,s$^{-1}$\fi}
\def\mujybm{\ifmmode{\rm \mu Jy}\,{\rm beam}^{-1}\else${\rm \mu}$Jy\,beam$^{-1}$\fi}
\def\ltsim{\ifmmode\stackrel{<}{_{\sim}}\else$\stackrel{<}{_{\sim}}$\fi}
\def\gtsim{\ifmmode\stackrel{>}{_{\sim}}\else$\stackrel{>}{_{\sim}}$\fi}
\def\farcs{\hbox{$.\!\!^{\prime\prime}$}}
\def\farms{\hbox{$.\!\!^{\prime}$}}
\def\fsec{\hbox{$.\!\!^{\rm s}$}}
\def\hour{\hbox{$^{\rm h}$}}
\def\min{\hbox{$^{\rm m}$}}
\def\mum{$\mu$m}
\def\spitzer{{\it Spitzer}}
\def\q24{q$_{\rm 24}$}

\begin{document}  
\title[An evolution of the IR-Radio correlation?]{An evolution of the IR-Radio
correlation at very low flux densities?}

\author[Beswick {\it et
al.}]{R.\,J.\,Beswick,$^1$\thanks{Robert.Beswick@manchester.ac.uk}
T.\,W.\,B.\,Muxlow,$\!^1$ H.\,Thrall,$\!^2$
A.\,M.\,S.\,Richards,$\!^1$ S.\,T.\,Garrington\,$^1$\\
$^1$MERLIN/VLBI National Facility, Jodrell Bank Observatory, The University of 
Manchester, Macclesfield, Cheshire, SK11~9DL\\
$^2$Jodrell Bank Observatory, The University of Manchester, Macclesfield, 
Cheshire, SK11~9DL }
 
\date{Accepted 2008 January 6.  Received 2008 January 3; in original form 2007 November 6}
\pagerange{\pageref{firstpage}--\pageref{lastpage}} \pubyear{2007}

\maketitle
\label{firstpage}
 
\begin{abstract} {In this paper we 
investigate the radio-MIR correlation at very low flux densities using extremely deep 1.4\,GHz sub-arcsecond angular resolution 
MERLIN$+$VLA observations of a 8\farms5$\times$8\farms5 field centred upon 
the Hubble Deep Field North, in conjunction with \spitzer\ 24\,\mum\ data.  From these results the MIR-radio correlation is extended to the very faint ($\sim$microJy) radio 
source population. Tentatively we detect a small deviation from the correlation 
at the faintest IR flux densities. We suggest that this small observed change in 
the gradient of the correlation is the result of a suppression of the MIR 
emission in faint star-forming galaxies.  This deviation potentially has 
significant implications for using either the MIR or non-thermal radio emission 
as a star-formation tracer of very low luminosity galaxies.}
\end{abstract}


\section{Introduction}

Since 1970s and 1980s studies of the radio and far-infrared (FIR) properties of 
galaxies have shown there to be a tight correlation between their emission in 
these two observing bands which extends over several orders of magnitude in 
luminosity \citep{vanderkruit73,condon82}. The advent of the {\it Infrared 
Astronomical Satellite} (IRAS) All Sky Survey in 1983 
\citep{neugebauer84,soifer87} enabled much larger systematic samples of galaxies 
to be studied at infrared wavelengths and hence further demonstrated the 
consistency and tightness of this correlation, albeit for relatively nearby 
sources \citep{helou85,dejong85,condon86,condon91,yun01}. Following {\it IRAS}, 
deep observations using the {\it Infrared Space Observatory} ({\it ISO})  
allowed fainter and higher redshift galaxies to be observed at mid-infrared (MIR) 
wavelengths. These {\it ISO} observations showed that the MIR emission 
from galaxies is loosely correlated with radio emission across a wide range of 
redshifts, tentatively extending to z$\sim$4 \citep{cohen00,elbaz02,garrett02}.

More recently the launch of the \spitzer\ Space Telescope in August 2003 
has greatly increased the sensitivity of MIR observations and hence our ability 
to study the MIR-radio correlation.  
Early results, such as from the \spitzer\ First Look Survey (FLS), have confirmed 
that the MIR-radio correlation holds for relatively bright star-forming galaxies (S$_{20\,{\rm cm}}>115$\,$\mu$Jy) out to at least a redshift of 1 
\citep{appleton04}.

Radio and infrared emission from galaxies in both the nearby and distant Universe is thought to arise from processes related to star-formation, hence resulting in the correlation between these two observing bands. The infrared emission is produced from dust heated by 
photons from young stars and the radio emission predominately arises from 
synchrotron radiation produced by the acceleration of charged particles from 
supernovae explosions. It has however recently been suggested that at low flux 
density and luminosities there may be some deviation from the tight well-known 
radio-IR correlation seen for brighter galaxies \citep{bell03,boyle07}.

\begin{center}
 \begin{figure*}
\setlength{\unitlength}{.5in}
\begin{picture}(16,5)
\put(-0.5, 0){\includegraphics{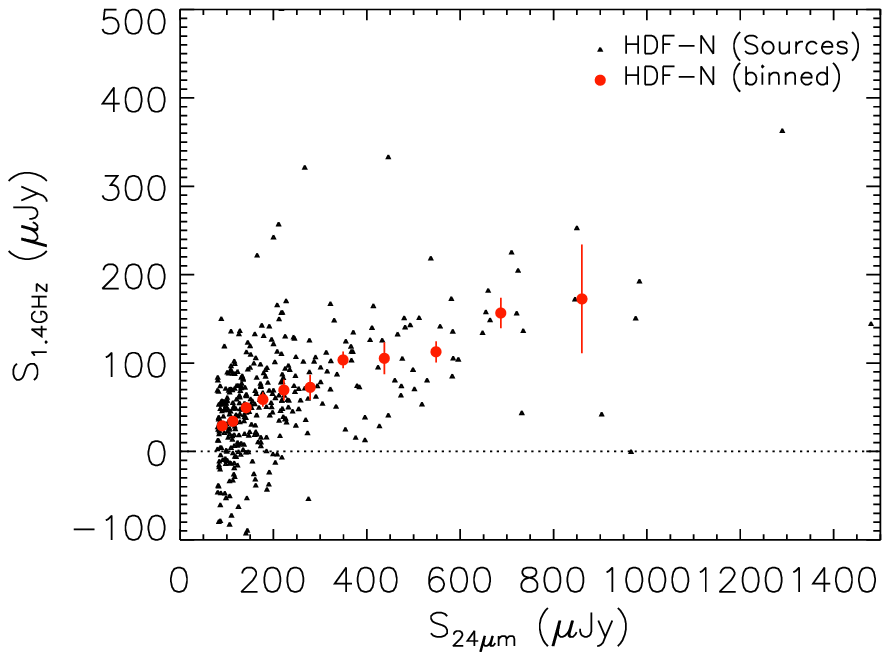}}
\put(6.5, 0){\includegraphics{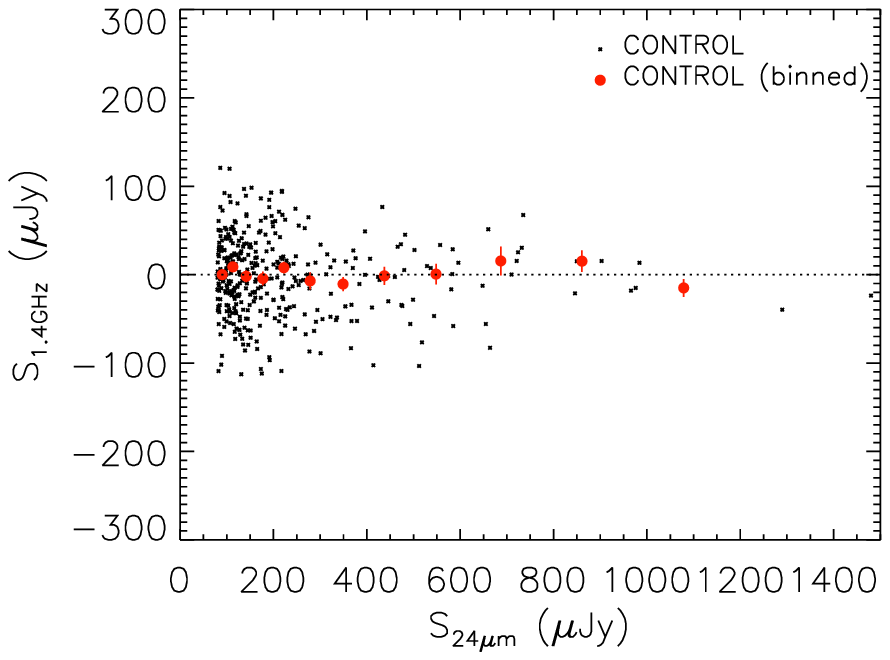}}
\end{picture}
\vskip -0.5cm
\caption{{\it Left-hand panel:} Radio 1.4\,GHz versus the MIR 24\,\mum\ flux 
density of all 377 individual sources (small triangle), and median radio flux 
density logarithmically binned by their 24\,\mum\ flux density (filled
circles) 
within the 8\farms5$\times$8\farms5 field. {\it Right-hand panel:} Control plot 
of 1.4\,GHz radio flux densities plotted against the 24\,\mum\ source flux 
densities of the sample. The radio flux densities of the control sample have 
derived in exactly the same manner as the flux densities plotted in the left-hand, 
panel but have been measured at randomly assigned sky positions in the 
8\farms5$\times$8\farms5 radio image where no known sources at any wavelength 
are located.}
\label{F20vsF24}
 \end{figure*}
\end{center}

\cite{bell03} argue that while the IR emission from luminous galaxies will trace 
the majority of the star-formation in these sources, in low luminosity galaxies the IR emission will be less luminous than expected considering the rate of star-formation within the source (i.e. the IR emission will not fully trace the star-formation). In this scenario the reduced efficiency of IR production relative 
to the source star-formation rate (SFR) would be the result of inherently lower 
dust opacities in lower luminosity sources and consequently less efficient 
reprocessing of UV photons from hot young stars into IR emission. The simple 
consequence of this is that at lower luminosities the near linear radio-IR 
correlation L$_{\rm radio}\propto$L$_{\rm IR}^{\!\gamma}$, with $\gamma >1$ 
\citep[e.g][]{cox88,price92} will be deviated from. {\it Of course such an 
assertion is dependent upon the radio emission providing a reliable tracer of 
star-formation at low luminosities which may be equally invalid.}

Recently \cite{boyle07} have presented a statistical analysis of Australia 
Telescope Compact Array (ATCA) 20\,cm observations of the 24\,\mum\ sources 
within two regions (the {\it Chandra} Deep Field South (CDFS) and the European 
Large Area {\it ISO} Survey S1 (ELAIS)) of the \spitzer\ Wide Field Survey 
(SWIRE). In this work \cite{boyle07} have co-added sensitive 
(rms$\sim$30\,$\mu$Jy) radio data at the locations of several thousand 24\,\mum\ 
sources. Using this method they have statistically detected the microJy radio 
counterparts of faint 24\,\mum\ sources. At low flux densities (S$_{\rm 24\, \mu 
m }=100\,\mu$Jy) they confirm the IR-radio correlation but find it to have a 
lower coefficient (S$_{\rm 1.4\,GHz}$\,=\,0.039\,S$_{\rm 24\,\mu m}$) than had 
previously been reported at higher flux densities. This coefficient is 
significantly different from results previously derived from detections of 
individual objects (e.g. \citealt{appleton04}) and is speculated by 
\cite{boyle07} to be the result of a change in the slope of the radio-IR 
correlation at low flux densities. 

In this paper, we utilise very deep, high resolution 20\,cm observations of the 
Hubble Deep Field North and surrounding area made using MERLIN and the VLA 
\citep{muxlow05} in combination with publicly available 24\,\mum\ \spitzer\ 
source catalogues from GOODS to study the MIR-Radio correlation for microjansky 
radio sources. This study extends the flux density limits of the radio-IR 
correlation by more than an order of magnitude for individual sources and 
overlaps the flux density regime studied using statistical stacking methods by 
\cite{boyle07}. Additionally we employ statistical stacking methods, similar to those 
used by \cite{boyle07}, to extend the correlation further to still lower flux 
densities.

We adopt H$_0=$75\,km\,s$^{-1}$\,Mpc$^{-1}$, $\Omega_{\rm m}=0.3$
and $\Omega_{\rm \Lambda}$=0.7 throughout this paper.

\section{Data and Analysis}
\subsection{MERLIN$+$VLA observations}

Extremely deep radio observations of the HDF-N region were made in 1996-97 at 
1.4\,GHz using both MERLIN and the VLA. These observations were initially 
presented in \cite{muxlow05}, \cite{richards98} and \cite{richards00}. The 
results from the combined 18 day MERLIN and 42\,hr VLA observations are described 
in detail in \cite{muxlow05}. The combined MERLIN$+$VLA image has an rms noise level of 
3.3\,$\mu$Jy per 0\farcs2 circular beam making it amongst the most sensitive, high-resolution radio images made to date.

Using the same methods as described in \cite{muxlow05} these combined MERLIN and 
VLA observations have been used to image the entire unaberrated field of view, 
8.5$\times$8.5 arcmin$^2$ in size, centred upon the original MERLIN pointing 
position ($\alpha=12$\hour\,36\min\,49\fsec4000, 
$\delta=+62$\degr\,12\arcmin\,58\farcs000 (J2000))\footnote{Public access to these radio data will be made 
available via an on-demand radio imaging tool developed by the authors using 
Virtual Observatory tools. A detailed explanation of this service can be found in 
\cite{richards07}.}. This image has an rms noise 
level of 3.6\,$\mu$Jy\,beam$^{-1}$ and has been convolved with a 0\farcs4 
circular beam. These observations have been shown to align with the International 
Coordinate Reference Frame (ICRF) to better than 15\,mas 
\citep{muxlow05}.

\subsection{GOODS-N \spitzer\ 24\,$\mu$m observations}

As part of the GOODS enhanced data 
release\footnote{http://www.stsci.edu/science/goods/DataProducts/} (DR1+ February 
2005) a catalogue of \spitzer\ 24\,$\mu$m source positions and flux densities for 
the GOODS-N field were released. This source catalogue is 
limited to flux densities $>$80\,$\mu$Jy providing a highly complete and reliable 
sample. At the time of writing this 24\,$\mu$m source catalogue represents the 
most complete and accurate mid-infrared source list publicly available for the 
GOODS-N/HDF-N field.

All 24\,\mum\ sources which are detected optically in GOODS {\it HST} ACS images 
show an accurate astrometric alignment with their optical counterparts implying 
that the astrometry between these two data-sets and their subsequent catalogues 
is self-consistent. However, a comparison of the astrometric alignment of the 
positions of sources catalogued by GOODS derived from their {\it HST} ACS images 
(Richards {\it et al.} 2007; Muxlow {\it et al.} {\it in prep}) shows there to be a systematic offset in 
declination of $-$0\farcs342 from the radio reference frame.  This linear 
declination correction, although small relative to the \spitzer\ resolution at 
24\,\mum\, is significant when compared to these high resolution radio data. This 
linear correction has been applied to the \spitzer\ source positions prior to all 
comparisons between the two data sets.

\begin{figure}
\begin{center}
\includegraphics[width=9cm]{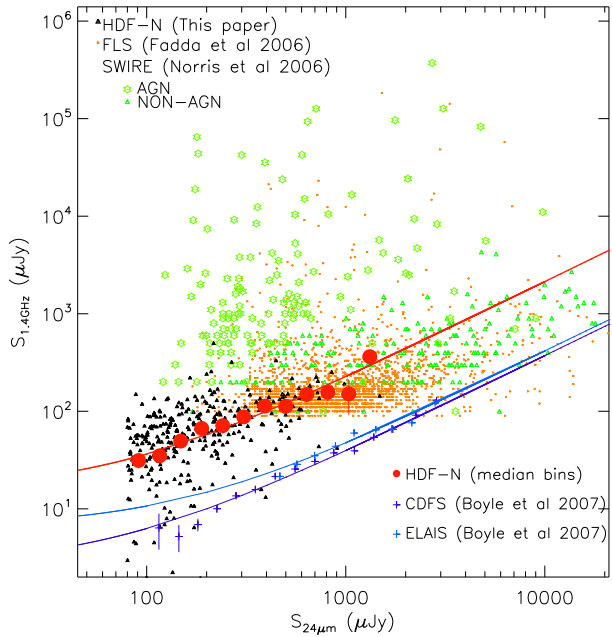}
\caption{Radio 1.4\,GHz versus the 24\,\mum\ flux density. Sources from the 
8\farms5$\times$8\farms5 HDF-N field are plotted individually (small black 
triangles). The median radio flux density logarithmically binned by 24\,\mum\ 
flux densities are plotted as large filled red circles. The solid red line represents the best-fit line to 
the binned HDF-N data alone. Sources within the CDFS-SWIRE field detected at both 
24\,\mum\ and 1.4\,GHz from Norris {\it et al.} (2006) are plotted as either 
green open stars (identified as AGN) or green triangles (not identified as AGN). 
All sources detected at both 1.4\,GHz and 24\,\mum\ in the \spitzer\ First Look 
Survey (FLS) with source position separations of $<$1\farcs5 are plotted in 
orange \citep{fadda06}.  Note the quantisation of the SWIRE and FLS points, in 
this and subsequent plots is a result of the accuracy of the flux densities 
tabulated in the literature. The blue pluses and fitted lines in the low portion 
of the plot show the IR-radio correlation derived from stacking radio emission at the positions of 24\,\mum\ 
sources in the CDFS and ELAIS field by Boyle {\it et al.} (2007). A
brief summary of the basic characteristics of each of the data sets
plotted included in Table\,\ref{tab2}.}
\label{F20vsF24log}       
\end{center}
\end{figure}

\begin{table*}
\centering
\caption{Summary of deep 1.4\,GHz radio and 24\,\mum\ surveys refereed to in this 
paper.}
\label{tab2}       
\begin{tabular}{cccccc}

Survey&Radio&Instrument&Radio senitivity&Angular&Reference\\
&Survey Area&&($\mu$Jy\,beam$^{-1}$)&Resolution&\\
\hline
HDF-N/GOODS-N&72.25\,arcmin$^2$&VLA+MERLIN&3.3&0\farcs2$\times$0\farcs2&Muxlow {\it et al.} 2005, this paper\\
ATLAS (CDFS/ELAIS)&3.7\,deg$^2$&ATCA&20$\rightarrow$60&11$\arcsec\times$5$\arcsec$&Norris {\it et al.} 2006\\
\spitzer\ FLS&5\,deg$^2$&VLA&23&5$\arcsec$&Condon {\it et al.} 2003, Fadda {\it et al.} 2006\\

\noalign{\smallskip}\hline
\end{tabular}
\end{table*}

\subsection{Measurement of source radio flux densities}

The field covered by these GOODS-N \spitzer\ 24\,\mum\ observations contains 1199 
24-\mum\ sources identified with flux densities $>$80\,$\mu$Jy.  The 
8.5$\times$8.5 arcmin$^2$ radio field contains 377 of these
\spitzer\ sources with 24\,\mum\ flux densities ranging from 80.1 to 1480\,$\mu$Jy.

Radio emission with the deep MERLIN+VLA data has been measured at the corrected 
position of each individual 24\,\mum\ source within the radio image. The radio 
flux density was measured within a series of concentric rings with a maximum 
radius of 4\,arcsec centred on each 24\,\mum\ source. Statistical analysis of these 
measurements and a sample of stronger radio sources within the same field shows 
that the radio flux density increases within progressively larger annuli out to a 
radius of 1\farcs5.  For stronger ($>5\sigma$) radio sources the total flux 
density recovered using this method within a radius of 1\farcs5 is 
between 90$\%$ and 100$\%$ of the source's total flux density as 
measured by other means, such as Gaussian fitting. Additionally the average 
(either mean or median) radial profile of all of the 24\,\mum\ sources shows that 
almost all of the radio flux density is recovered within radii less than 1\farcs5 
(\citealt{beswick06}; Beswick {\it et al.} {\it in prep}). Throughout the rest of 
this paper the radio flux density recovered within a radius of 1\farcs5 of each 
IR \spitzer\ source will be assumed to be equal to the total radio source flux density.

\begin{center}
 \begin{figure*}
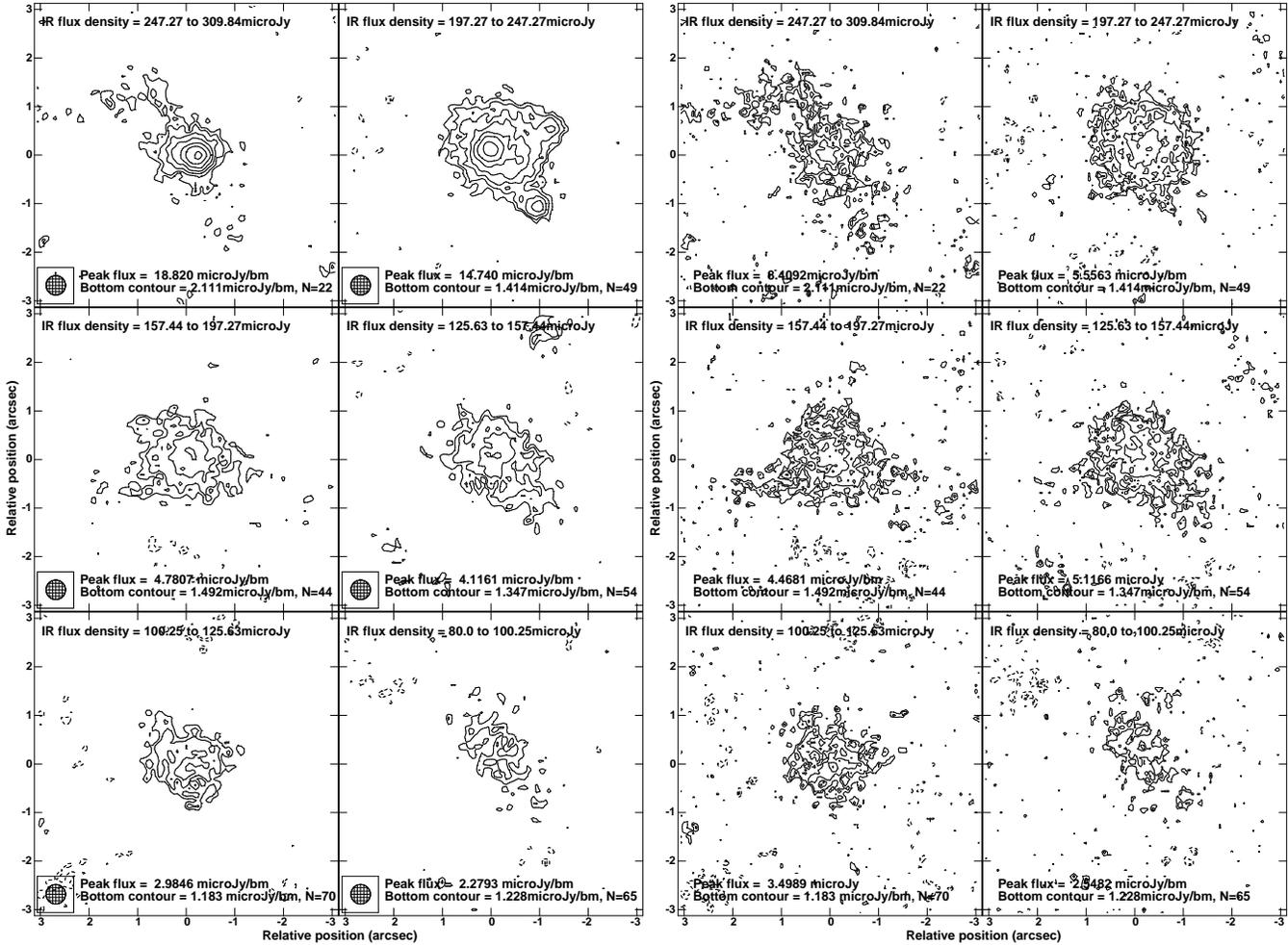

\setlength{\unitlength}{.5in}
\begin{picture}(16,11)
\put(-1.2, 0){\includegraphics{Beswick-fig3a.ps}}
\put(5.8, 0){\includegraphics{Beswick-fig3b.ps}}
\end{picture}
\vskip -1cm
\caption{Mean (left) and median (right) images of the 1.4\,GHz radio emission 
for all sources within the six faintest 24\,\mum\ flux density logarithmic bins plotted 
in Fig.\,1 in descending flux density order from top-left to bottom-right. The 
range of 24\,\mum\ flux density over which each image has been stacked is shown 
at the top of individual panels. Each image is contoured with levels of $-$2, 
$-$1.414, $-$1,1, 1.414, 2, 2.828, 4, 5.657, 8, 11.31, 16, 22.63 and 32 times 
3$\times$(3.3/$\sqrt {\rm N}$)\,$\mu$Jy\,bm$^{-1}$, where N equals the number of 
24\,\mum\ source positions averaged in the map. The peak flux density, lowest 
plotted contour and number of IR sources which have been averaged over (N) is 
listed at the bottom of each image panel.}
\label{maps}
 \end{figure*}
\end{center}

As a confirmation of this method several thousand random sky 
positions within the radio field were selected and cross-referenced to
exclude any which were at 
the positions of known sources at any wavelengths. At each of these blank 
positions the radio flux density was measured using an identical method as 
outlined above. The flux densities recorded within each of these annuli showed no 
positive bias within any ring and followed the same Gaussian distribution as 
derived from the pixel-by-pixel noise statistics of the whole image once all 
radio sources greater than 40\,$\mu$Jy have been
subtracted. As further test of these methods 90 fake,
faint ($<3\sigma$) and partially-resolved radio sources were injected into the {\it u-v} data. These data were
then Fourier inverted and cleaned in an identical manner to the real
data and then the flux densities of these fake sources were extracted
using the methods outlined above.  Whilst, as expected, these sources were not detected
individually their radio emission, when averaged together, was
consistent with the average flux density of the
population of fake sources injected into the data.

A more detailed description of this method and the radio characteristics of this 
sample of 24\,\mum\ \spitzer\ source, including source catalogues and size 
information, will be presented in a forthcoming paper.

\section{Results}

Of the 377 24\,\mum\ \spitzer\ sources within the radio field 303 were 
found, by this method, to have total radio flux densities greater than 3 times 
the local map rms. Many of these source have peak flux densities below 3-$\sigma$ 
and hence were not selected using the radio-only detection methods employed by 
\cite{muxlow05}. The 1.4\,GHz versus 24\,\mum\ flux densities of all 377 of 
these source are plotted in left-hand panel of
Fig\,\ref{F20vsF24}. The control 
sample, plotted in the right-hand panel of Fig\,\ref{F20vsF24}, is
derived using an identical radio flux extraction methods and from the same radio data 
but at randomly assigned locations coincident with no known source position.  The flux densities of these sources are also plotted in 
Fig\,\ref{F20vsF24log}, along with sources detected in the shallower surveys of the CDFS-SWIRE field by \cite{norris06} using ATCA, and 
the \spitzer\ FLS using the VLA by \cite{fadda06}. Sources from the southern 
CDFS-SWIRE field which have been categorised by \cite{norris06} as containing AGN 
(plotted as stars) are generally situated toward the upper half of this figure 
and show an expected radio excess when compared to their 24\,\mum\ flux 
densities. Source which were not identified as AGN by \cite{norris06} are plotted 
as triangles in this figure. It should be noted that these sources have not been classified as containing an AGN but may in fact be either AGN, 
star-forming galaxies or a combination of the two.

Overlaid in blue in the lower half of Fig.\,\ref{F20vsF24log} are the flux densities of 
groups of median stacked infrared sources from the ELAIS and CDFS field from 
\cite{boyle07} along with fits to these data. Also overlaid, as filled
red circles, in this
figure are the median values of all of the HDF-N sources logarithmically binned by 
their 24\,\mum\ flux densities. There is a clear discrepancy between
the averaged radio flux densities derived from these HDF-N
observations and the flux densities from the ELAIS and CDFS fields \citep{boyle07} 
which will be discussed further in the following sections.

\begin{table*}
\centering
\caption{Flux measurements from stacked images and binned data. The flux density 
in the stacked images has been measured by fitting a single Gaussian to the radio 
emission. The errors attached to measurements made from the stacked images are 
the formal errors of these Gaussian fits, rather than the image rms which in each 
case is comparable to $\sigma$. Sources within 3\,arcsec of the edge of the radio 
field have been excluded from the stacked images, hence slightly reducing the 
number of sources combined in each image relative to those included in the binned 
data.}
\label{tab1}       
\begin{tabular}{rrrrrrrr}
\hline\noalign{\smallskip}
 &  &   \multicolumn{3}{c}{Binned data}
&\multicolumn{3}{c}{Stacked images}\\
&&Median&Mean&&&Median&Mean\\
S$_{24\,{\rm \mu m}}$ & Number &S$_{\rm 1.4\,GHz}$&S$_{\rm 1.4\,GHz}$&$\sigma_{\rm 1.4\,GHz}$&Number&S$_{\rm 1.4\,GHz}$&S$_{\rm 1.4\,GHz}$  \\
($\mu$Jy)&  &($\mu$Jy)&($\mu$Jy)&($\mu$Jy)&&($\mu$Jy)&($\mu$Jy)\\
\hline
88.8&65&28.9&22.6&0.68&65&26.7$\pm$8&27.8$\pm$6\\
112.5&72&34.1&32.0&0.62&70&34.0$\pm$7&32.9$\pm$5\\
138.9&57&49.5&47.7&0.81&54&63.1$\pm$9 &58.1$\pm$7\\
174.7&46&58.8&56.1&1.21&44&64.8$\pm$10&60.4$\pm$8\\
219.4&49&69.4&90.7&1.68&49&76.1$\pm$8 &90.4$\pm$6\\
275.0&22&72.5&79.0&3.01&22&81.9$\pm$13&53.3$\pm$4\\
\noalign{\smallskip}\hline
\end{tabular}
\end{table*}

Using an analysis method similar to \cite{boyle07} the mean and median images of 
the radio emission from HDF-N field of the 6 lowest 24\,\mum\ flux 
density logarithmic bins (as plotted in Figs.\,\ref{F20vsF24} and 
\ref{F20vsF24log}) have been derived and are shown in Fig.\,\ref{maps}. Each of 
these images is the statistically combined radio emission from the location of 
several tens of \spitzer\ 24\,\mum\ sources and has been contoured at multiples 
of three times 3.3\,$\mu$Jy\,beam$^{-1}$ 
divided by $\sqrt{\rm N}$ where N equals the number of \spitzer\ source positions 
that have been stacked together. As can be seen in these images the off-source 
noise levels achieved approaches the value expected when co-adding multiple images 
with near-Gaussian noise properties. The co-added image rms achieved in the faintest 24\,\mum\ flux density bin (80.0 to 100.25\,$\mu$Jy) is 
0.45 and 0.56\,$\mu$Jy\,beam$^{-1}$ in the mean and median 
co-added images respectively. The flux densities measured directly from these 
stacked radio images are consistent with the statistical mean and median flux 
densities within the equivalent bins derived from the individual source flux 
extraction method described in Section 2.3 (Table\,\ref{tab1}) and plotted in 
Figs.\,\ref{F20vsF24} and \ref{F20vsF24log}. In each stacked image the largest angular size of the 
regions showing elevated radio emission within Fig.\,\ref{maps} is 
enclosed within the 1\farcs5 radius circular area from which the individual 
source flux densities have been extracted.

The sizes of the composite radio sources in these co-added images 
(Fig.\,\ref{maps}) represent a combination of the true radio source size, any 
offsets between the IR and radio emitting regions within the sources and any 
random errors in the \spitzer\ source positions, all of which will be convolved 
with the synthesised beam applied to these radio data (0\farcs4
circular beam in this case). No 
systematic offset is seen between the emission in these composite radio images 
from their nominal image centres implying that no significant non-random errors 
between the astrometric alignment of these data-sets remain. Further analysis of 
the radio source sizes and structures will be the subject of a future paper.

In Fig.\,\ref{maps} the mean (left) images of the radio emission for the two 
highest 24\,\mum\ flux density bins are dominated by emission from a few bright 
sources. Several of these brighter radio sources can be seen in 
Fig.\,\ref{F20vsF24} scattered above S$_{\rm 1.4\,GHz}=$200\,$\mu$Jy. The 
presence of these brighter radio sources and the low numbers of images stacked 
together is clearly evident when the mean and median averaged images are 
compared. In particular in the second brightest bin (197.27 to 247.27\,$\mu$Jy) 
two mJy radio sources are present and offset from their equivalent 24\,\mum\ 
position. It is likely that these bright radio sources, which deviate from the 
radio-IR correlation, may contain radio AGN which would feasibly be spatially 
offset from any ongoing star-formation and 24\,\mum\ position.

\begin{figure}
\begin{center}
\includegraphics[width=9cm]{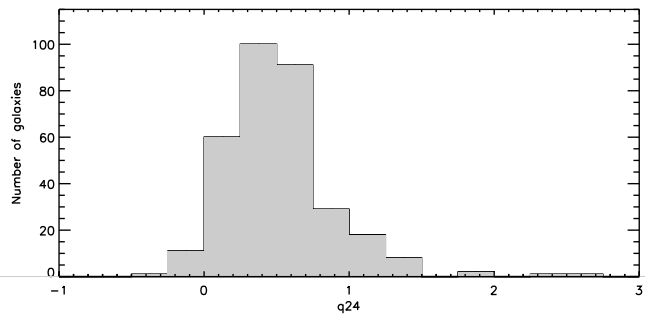}
\caption{Histogram of the distribution of \q24\ for the 377 sources within the 8\farms5$\times$8\farms5 field 
centred upon the HDF-N.}
\label{histq24}
\end{center}
\end{figure}

An alternative method of projecting the correlation between these 24\,\mum\ and 
1.4\,GHz data is as a function of the commonly defined parameter q$_{\rm 24}$, 
where q$_{\rm 24}$ = log(S$_{\rm 24\,\mu m}$/S$_{\rm 1.4\,GHz}$).  The mean and 
median value of \q24\ for all of the HDF-N sources presented in this study is 
0.52 and 0.48 respectively, compared to the value of 0.69$\pm$0.36 and 0.73 
derived from non-AGN sources catalogued in observations of the SWIRE field by 
\cite{norris06}.  A similar study of brighter sources in the \spitzer\ FLS 
\citep{appleton04} show the mean value to be (\q24\,=0.84$\pm$0.28) slightly 
higher than average value derived here. The distribution of \q24\ values for all 
sources in this HDF-N study is shown in Fig.\,\ref{histq24}.

In Figs.\,\ref{q24_1}, \ref{q24_1b}, \ref{q24_2} and \ref{q24_3} the parameter \q24\ is plotted 
against S$_{\rm 24\,\mu m}$, S$_{\rm 1.4\,GHz}$ and redshift. The 
monochromatic value of \q24\ represents the slope of the IR-radio correlation 
and the size of dispersion in \q24\ is related to the strength of the 
correlation. In each of these figures no {\it k}-correction has been applied to either 
the radio or IR luminosities since we have incomplete redshifts for the sample, 
and inadequate spectral information is known about the majority of the
individual sources in 
either the IR or radio bands. A similar study by \cite{appleton04} has shown 
that {\it k}-corrections made to individual sources result in only a minimal systematic 
effect on the average value of \q24\ below redshift $\sim$1. 

 In Fig.\,\ref{q24_1} the values of \q24\ for sources from these HDF-N data, the 
SWIRE field \citep{norris06} and the \spitzer\ FLS field \citep{fadda06} 
are plotted against 24\,\mum\ flux density. In the upper panel both the AGN 
identified and non-AGN sources of \cite{norris06} are shown whilst in the lower 
panel the AGN sources are excluded for clarity. In the upper panel the AGN 
sources from the SWIRE field, as expected, deviate from the correlation generally 
showing an excess of 1.4\,GHz radio emission compared to their 24\,\mum\ 
emission. The HDF-N field covers a considerably smaller area and was originally 
chosen to have few bright radio-AGN, which is confirmed by the minimal number of 
HDF-N sources displaying an excess of 1.4\,GHz radio emission.  Amongst the 
HDF-N sources and the non-AGN sources from the SWIRE field \citep[and 
Fig.\,\ref{q24_1} here]{norris06} \q24\ shows a small trend toward lower values 
at smaller 24\,\mum\ flux densities. The SWIRE sources plotted here have been selected to 
have both radio and IR flux density above a certain threshold. The lowest radio 
flux density of sources within the SWIRE field catalogue \citep{norris06} is 
100\,$\mu$Jy. A line of constant radio flux density (100\,$\mu$Jy, the lowest individual source detections in the SWIRE catelogue) is overlaid on 
Fig.\,\ref{q24_1} above which sources from the SWIRE study are excluded. The HDF-N radio data plotted here has a sensitivity limit $\sim$10 times lower than that of the SWIRE data, thus potentially only 
excluding sources in the extreme top left-hand part of this plot. Furthermore, 
the sample of sources plotted here has been subject to no radio detection 
threshold limit hence removing this potential bias completely.

The \q24\ values for the HDF-N data and the sources from
\cite{norris06} binned as a function of S$_{\rm 24\,\mu m}$, along
with the ratio of S$_{\rm 24\,\mu m}$ and S$_{\rm 1.4\,GHz}$ are shown
in Fig\,\ref{q24_1b}.  Especially within the binned points, these two
plots show a tentative trend for low \q24\ and  $\frac{{\rm S}_{\rm
24\,\mu m}}{{\rm S}_{\rm 1.4\,GHz}}$ with declining 24\,\mum\ flux
density. The gray areas and dotted line in these two
panels show the value of the \q24\ as derived from the \spitzer\ FLS
\citep{appleton04}. The HDF-N observations presented here are
consistent with those of \cite{appleton04} at high 24\mum\ flux
densities. 

The effect of the 24\,\mum\ sensitivity cut-off (80.1\,$\mu$Jy) in the GOODS-N 
data and the equivalent limit in the data of \cite{norris06} will, however, 
positively bias measurements of \q24\ at low radio flux densities. This is 
clearly shown in Fig.\,\ref{q24_2}. The lower flux density limit imposed by the 
\spitzer\ sensitivity results in an exclusion of sources in the lower portion of 
Fig.\,\ref{q24_2}. This consequently biases the determination of the value of 
\q24\ for any complete sample of sources with low radio flux densities.

Of the 377 \spitzer\ source within the 8\farms5$\times$8\farms5 HDF-N radio 
field, 259 of these galaxies have published spectroscopic or photometric 
redshifts. The \q24\ values for this subset of sources, along with 50 non-AGN 
SWIRE sources, are plotted with respect to their redshift in Fig\,\ref{q24_3}. 
Within these two data-sets \q24\ is seen to slightly reduce at higher 
redshifts although this effect is small and significantly less than the scatter. 
These values of \q24\ have not been {\it k}-corrected. The 
application of a {\it k}-correction to these data will result in a small increase in 
\q24\ which will increase as a function of z.  It is interesting to also note 
that recent studies of discrete areas within 4 very nearby star-forming galaxies 
\citep{murphy06a} have shown \q24\ to vary across the extent of individual 
sources. However, \cite{murphy06a} derive a mean value of \q24$\approx$1 when 
integrating over entire galaxies in their sample, at z$\approx$0, which is 
consistent with many of the source in both the SWIRE and HDF-N samples but 
somewhat higher than the average value derived at higher redshifts.

Using the available redshifts the observed luminosity has been calculated and is 
plotted in Fig.\,\ref{lumlum}, including both the sources in the HDF-N region and the 
non-AGN sources in the SWIRE field.  As 
can be seen in Fig.\,\ref{lumlum} the 24\,\mum-to-1.4\,GHz correlation, for 
individual sources, extends down to L$_{\rm 
1.4\,GHz}\approx10^{21}$\,W\,Hz$^{-1}$ (L$_{\rm 24\,\mu 
m}\approx10^{21.7}$\,W\,Hz$^{-1}$).

\section{Discussion}

\subsection{Radio emission from faint 24\,\mum\ sources:-\\ Extending the 
IR-Radio correlation}

The primary results of this work demonstrate that this sample of individual 
24\,\mum\ selected sources follow the MIR-radio correlation down to radio flux 
densities of a few microJy. These results are consistent with shallower 
independent radio surveys (such as the ATCA observations of the SWIRE
field) and seamlessly extend the IR-radio correlation down to lower flux densities (see 
Fig.\,\ref{F20vsF24log}).

The majority of the IR selected sources within the HDF-N region have radio flux 
densities less than 100\,$\mu$Jy, with only a few sources which show a 
significant excess of 1.4\,GHz flux density, when compared to the IR-radio 
correlation. This implies that the vast majority of these IR-selected microJy 
radio sources are primarily driven by star-formation with little significant 
contamination from AGN.  This is consistent with the results of \cite{muxlow05} who find that 
below 100\,$\mu$Jy greater than 70 percent of radio sources are starburst 
systems. However, the data presented here are based on an IR
selected sample rather than a radio selected sample. This will
inherently result in the sample being dominated by star-forming
systems but will not exclude the presence of a population of weak microJy AGN sources.

At the positions of the majority of the lowest 24\,\mum\ flux densities sources 
the peak radio emission is not significantly greater than the image rms. As a 
consequence most of these sources were not formally identified in the previous 
radio study \citep{muxlow05}. In many case, especially at low 24\,\mum\ flux 
densities, the extracted total radio flux density of individual sources is 
close to the radio sensitivity of these data (Fig\,\ref{F20vsF24}). When directly 
compared with the control sample (Fig\,\ref{F20vsF24}, {\it Right}), which has 
been compiled by extracting the radio flux density from blank areas within the 
same data using identical methods, a significant correlation between the strength 
of the total radio and 24\,\mum\ emission is evident. However, the
radio sensitivity limit of these data precludes the detailed
discussion of individual low flux density sources.

\begin{figure}
\begin{center}
\includegraphics[width=9cm]{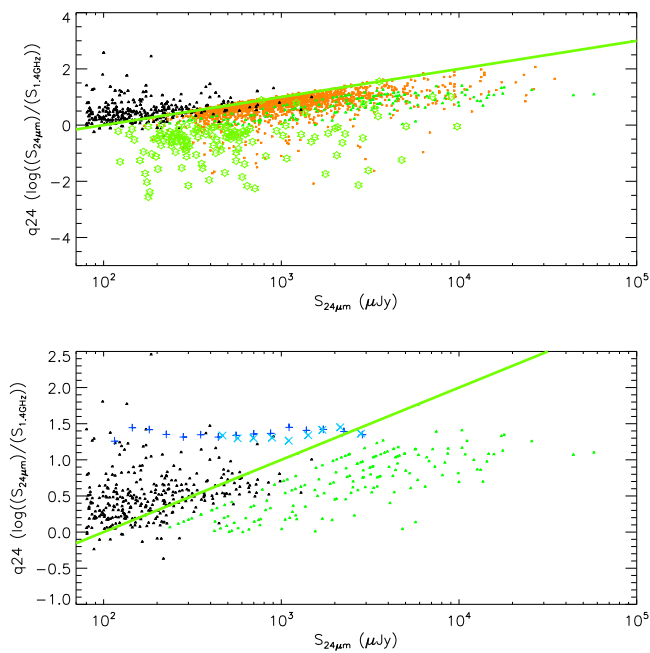}
\caption{The 24\,\mum\ flux density versus q$_{24}$. All 377 source from the 8\farms5$\times$8\farms5 HDF-N 
field are plotted as individual sources (small black triangle).  In the upper 
panel sources detected at both 24\,\mum\ and 1.4\,GHz in the SWIRE-CDFS field 
from Norris {\it et al.} (2006) are plotted in green (small triangles
are sources not identified as AGN and open stars have been identified 
as AGN). Sources from the \spitzer\ FLS are plotted in orange from Fadda {\it et 
al.} (2006). In the lower panel all of the \spitzer\ identified sources in the 
HDF-N and the only non-AGN sources from Norris {\it et al.} (2006) are plotted. 
Additionally plotted as `pluses' and 'crosses' in the lower panel are the values 
of q$_{24}$ as derived from the stacking analysis by Boyle {\it et al.} (2007) of 
the SWIRE-CDFS and ELAIS fields respectively. The solid green line overlaid on both 
panels is a line of constant radio flux density of 100\,$\mu$Jy, the
lowest radio flux 
density of sources in the SWIRE sample plotted here.}
\label{q24_1}       
\end{center}
\end{figure}

\begin{figure}
\begin{center}
\includegraphics[width=9cm]{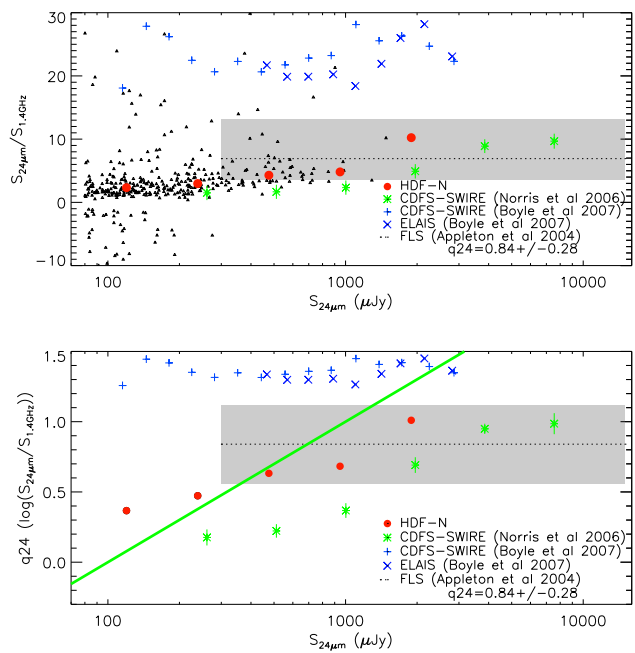}
\caption{In the upper panel the flux density ratio (${\rm S_{24\,\mu m}}/{\rm 
S_{1.4\,GHz}}$) versus 24\,\mum\ flux density is shown. The individual 
sources from the HDF-N field are plotted as small black triangles, the median values of these 
HDF-N source binned as a function of S$_{24\,\mu m}$ as filled circles, green 
stars show the median binned values for sources listed as non-AGN within the 
CDFS-SWIRE sample of Norris {\it et al.} (2006) and the blue `crosses' and `pluses' show the flux density ratios derived from the stacking analysis of the 
SWIRE-CDFS and ELAIS fields respectively from Boyle {\it et al.}
(2007). The overlaid 
black dotted line is the mean value for (${\rm S_{24\,\mu m}}/{\rm 
S_{1.4\,GHz}}$) derived by Appleton {\it et al.} (2004). This line is equivalent 
to q$_{24}$=0.84$\pm$0.28 with the gray filled box representing the area enclosed 
by these errors and the flux density range investigated by Appleton {\it et al.} 
(2004). The values of q$_{24}$ against 24\,\mum\ flux density are plotted in the 
lower panel.  The symbols within this plot are identical to those in the upper 
panel, individual HDF-N sources are not included for clarity. The additional 
diagonal solid green line represents a line of constant radio flux density of 
100\,$\mu$Jy, the lowest flux density of sources in the SWIRE sample plotted here 
(sources above this line are excluded by this limit from the SWIRE sample). This 
flux density limit will significantly effect the values of the binned SWIRE data 
points (green crosses) negatively biasing the four lowest flux density bins. This 
bias only effects the SWIRE sample. }
\label{q24_1b}       
\end{center}
\end{figure}

\begin{figure}
\begin{center}
\includegraphics[width=9cm]{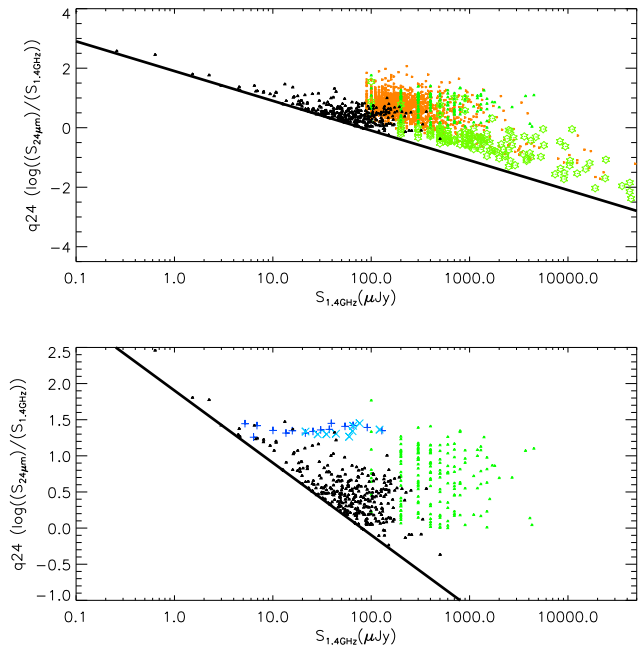}
\caption{The 1.4\,GHz flux density versus q$_{24}$. The samples
plotted and symbols used are the same as those plotted in Fig\,\ref{q24_1}.  In 
both plots the solid line represents a line of constant S$_{24\,\mu 
m}$=80.1\,$\mu$Jy which is equal to the detection threshold of the GOODS-N 
\spitzer\ sample used.}

\label{q24_2}       
\end{center}
\end{figure}

\subsubsection{The average radio emission from the faintest IR sources}

The radio emission from many of the individual low flux density 24\,\mum\ sources 
is too faint to be imaged with high signal-to-noise in these radio data. However, 
it is possible to characterise their average radio emission by either binning the 
measured radio excess at their locations or by forming composite radio images 
from many sources. With the relatively limited sample of IR sources which are 
co-spatial with these radio observations (377 in total) it is only statistically 
viable to use either binning or image stacking methods at the lowest IR flux 
densities where the number of sources that can be combined becomes large.

Overlaid in Figs.\,\ref{F20vsF24} and \ref{F20vsF24log} are the median flux 
densities of IR sources in the HDF-N region imaged here. These bins have been 
logarithmically sampled as a function of 24\,\mum\ flux density and include all 
sources within these bins regardless of their measured radio flux density. Whilst 
at the lowest emission levels the radio flux densities of individual
sources will be subject to moderate or large errors the binned ensemble of these
points will robustly represent the flux distribution of sample as a
whole. Using the radio flux densities extracted at the positions of the
individual 24\,\mum\ sources, significant radio emission is
detected in every averaged 24\,\mum\ flux density bin.

To further test the viability of measuring the 
statistical radio flux density from sources in this sample below the formal radio 
detection limit of these MERLIN+VLA data, the images have been stacked at the 
positions of multiple IR sources, binned as a function of their 24\,\mum\ flux 
density (Fig.\,\ref{maps}). These postage-stamp images have been averaged on a 
pixel-by-pixel basis, where each pixel in the stacked image is either the mean or median 
value of the pixels within the individual 24\,\mum\ source image stacks. Each of the 6 lowest bins, for both the mean and 
median averaged images, show significant levels of radio emission.  
The 1.4\,GHz flux density in these images, extracted via Gaussian 
fitting, is consistent with the binned level of emission derived from measuring 
individual sources, demonstrating the validity of either method 
(Table\,\ref{tab1}).

 The sub-arcsecond 
angular resolution of these MERLIN+VLA observations, in conjunction with their 
relatively good {\it u-v} coverage, allow the average radio structures of faint 
IR sources to be investigated.  The median and mean images in Fig.\,\ref{maps}, 
especially in the two highest flux density bins show some distinct 
differences. These differences arise due to inclusion of a few bright radio 
sources which dominate the average emission included in these bins. In particular in the second highest 
flux density mean image several `bright' radio sources are offset from the map 
centre. The positional offset of radio emission from these sources is real and 
indicates that, at least in these cases, that the origins of the radio and IR 
emission are different.

The Gaussian fitted sizes of the radio emission in the stacked images 
(Fig\,\ref{maps}) provide an upper limit on the average size of the radio 
counterparts of these faint IR sources. The largest angular sizes of the radio 
emission in the median stacked images created from this sample range between 
1\farcs4 and 2\arcsec. This is approximately equivalent to a linear size of 
10\,kpc at redshifts beyond 1. These upper limits on the radio source sizes are 
consistent with radio emission on galactic and sub-galactic scales and originating within kpc-scale starburst
systems. A more detailed analysis of the sizes and structures of the radio emission 
from both these composite images (Fig.\,\ref{maps}) and the individual sources 
themselves is beyond the scope of this paper and will be presented in a latter 
publication.

\begin{figure}
\begin{center}
\includegraphics[width=9cm]{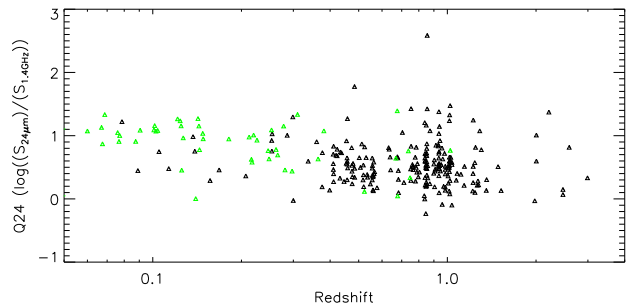}
\caption{The q$_{24}$ versus 
redshift. All 259 source from the 8\farms5$\times$8\farms5 HDF-N field are 
plotted as individual sources (black triangle) with known redshifts. Fifty 
sources from the CDFS-SWIRE field detected at both 24\,\mum\ and 1.4\,GHz, and 
redshift information from Norris {\it et al.} (2006) are plotted as green 
triangle. Only sources not identified as AGN by Norris {\it et al.} 2006 are 
plotted. No k-correction has been applied to the data points plotted.}
\label{q24_3}       
\end{center}
\end{figure}

\subsubsection{Comparison with other results}

Two directly comparable studies of the 1.4\,GHz radio and 24\,\mum\ MIR emission 
from sources in the \spitzer\ FLS and the SWIRE fields have been made by 
\cite{appleton04} and \cite{boyle07} respectively. \cite{appleton04} compared 
observations of the \spitzer\ FLS made by \cite{condon03} using the VLA in its 
B-configuration (5\arcsec restoring beam) with 24\,\mum\ and 70\,\mum\ \spitzer\ 
observations. These observations had a detection threshold of 90\,$\mu$Jy and 
500\,$\mu$Jy at 1.4\,GHz and 24\,\mum\ respectively. 
 Whereas \cite{boyle07} co-added deep ($\sigma \sim$30\,$\mu$Jy/bm) 
1.4\,GHz ATCA observations of the southern SWIRE field at the positions of many 
thousands of \spitzer\ 24\,\mum\ sources, regardless of their detection at radio 
wavelengths, in order to statistically detect radio emission at the few $\mu$Jy 
level.  As a comparison to the HDF-N observations presented here, data points 
from the SWIRE field \citep{norris06}, the \spitzer\ FLS \citep{fadda06} and the 
co-added flux densities derived by \cite{boyle07} are plotted in 
Figs.\,\ref{F20vsF24log}, \ref{q24_1}, \ref{q24_1b} and \ref{q24_2}.

Using the \spitzer\ FLS \cite{appleton04} studied MIR to radio correlation for 
508 sources with redshift information, confirming results from similar 
previous surveys. Following this work \cite{fadda06} have published a more 
comprehensive catalogue of $\sim$17,000 24\,\mum\ sources from the \spitzer\ FLS 
of which 2415 have radio counterparts within 4\arcsec. Just over 2000 sources with 
radio counterparts within 1\farcs5 are plotted in Figs.\,\ref{F20vsF24log}, 
\ref{q24_1} and \ref{q24_2}. In general the sources plotted follow the MIR to 
radio correlation but these data contain many outliers with elevated radio 
emission, characteristic of galaxies containing AGN.

The study of \cite{appleton04} included sources with S$_{24\mu m}\gtsim$0.5\,mJy 
and S$_{1.4\,{\rm GHz}}>$90\,$\mu$Jy. The majority of this sample have redshifts 
of $\ltsim$1.  The mean (non-{\it k}-corrected) value of
 q$_{24}$ derived by \cite{appleton04} is
0.84$\pm$0.28, which is shown in Fig\,\ref{q24_1b}. The non-AGN sources detected at both 1.4\,GHz and 24\,\mum\ within the
SWIRE field catalogued by \cite{norris06} are also plotted in
Figs.\,\ref{F20vsF24log}, \ref{q24_1} and \ref{q24_2}. These sources
are limited to $\geq$0.1\,mJy and $\gtsim$170$\mu$Jy at 1.4\,GHz
and 24\,\mum\ respectively. Using only the sources within the SWIRE field
identified as not being AGN by \cite{norris06} the mean
(non-{\it k}-corrected) values of q$_{24}$ is 0.69$\pm$0.39.  The radio flux
density threshold of both of these surveys overlaps with the brighter
radio sources in this MERLIN+VLA study. 

These deep HDF-N MERLIN+VLA observations trace individual sources with considerably fainter 
flux density than those observed in the SWIRE and \spitzer\ FLS surveys. These 
observations (Figs.\,\ref{F20vsF24log}, \ref{q24_1} and \ref{q24_2}) are 
consistent with the extrapolation to lower flux densities of results found 
previously \citep[e.g.][]{appleton04,norris06,fadda06}. However both the SWIRE and 
\spitzer\ FLS surveys which sample sources with somewhat higher flux densities 
sources have mean values of q$_{24}$ which are consistent but slightly higher 
than the value of 0.52$\pm$0.37 derived from these HDF-N data.

Averaging together the radio emission from multiple 24\,\mum\ \spitzer\ source 
within the CDFS and ELAIS fields \cite{boyle07} have 
traced the radio emission against MIR sources with flux densities comparable to 
those detected in these more sensitive HDF-N observations.  \cite{boyle07} have exploited the large number of MIR sources 
detected in these fields to average together multiple radio images made at the 
MIR source positions and statistically detect radio
emission significantly below their radio detection threshold.  This
method is directly comparable to the methods applied to these 
radio observations of the HDF-N and is identical to the methods used to 
produce the stacked images presented in Fig.\,\ref{maps} and Table\,\ref{tab1}. 
The median stacked radio flux densities of MIR sources derived by \cite{boyle07} show a 
significantly reduced S$_{\rm 1.4\,GHz}$ relative to S$_{24\,\mu {\rm m}}$ 
(i.e. higher q$_{24}$ value) compared with the distribution of individual sources detected 
in the \spitzer\ FLS \citep{fadda06,appleton04}, the catalogued SWIRE sources 
\citep{norris06} or the HDF-N results presented here.   Whilst there are several differences between the studies of \cite{boyle07} and those presented here, such 
as the sensitivity (rms\,$\sim$30\,$\mu$Jy versus $\sim$3\,$\mu$Jy) and angular 
resolution (6\arcsec versus 0\farcs4) of the two surveys, it remains hard to 
simply reconcile these results.

\cite{boyle07} recognised the inconsistency of their calculated values of 
q$_{24}$(=1.39) when compared with 0.84 derived by \cite{appleton04} and have 
extensively simulated and tested the correctness of their analysis methods. Since 
the results of \cite{boyle07} are based on an infrared-selected sample with no 
radio detection limit, this discrepancy could possibly be explained by the 
presence of a previously unknown population of faint radio sources with a low 
radio to infrared flux density ratio. However the values derived by \cite{boyle07} are not confirmed in the study presented here. 

A possible alternative explanation may be that the ATCA
observations are underestimating the radio flux density of sources at
the lowest flux densities when compared to these other surveys which
have used radio data obtained using the VLA and/or MERLIN.
It is interesting to note that an earlier
deep survey using the ATCA at 1400\,MHz (ATESP, Australia Telescope ESO Slice
Project) \cite{prandoni00} found, upon comparison with NVSS flux
densities for the same field, that whilst for bright sources the ATCA
and NVSS flux density values were consistent, for faint sources
the ATCA results underestimated the radio flux densities by up to a factor
of 2 compared with the VLA NVSS. This underestimation of radio flux density was seen by
\cite{prandoni00} to increase with diminishing radio brightness.
Whilst it remains unclear as to the cause, or reality, of this effect, if
applicable to the ELAIS and CDFS fields, it could result in the elevated
values of \q24\ seen by \cite{boyle07} at very low radio flux densities.

\begin{figure*} 
\begin{center} 
\includegraphics[width=16.0cm]{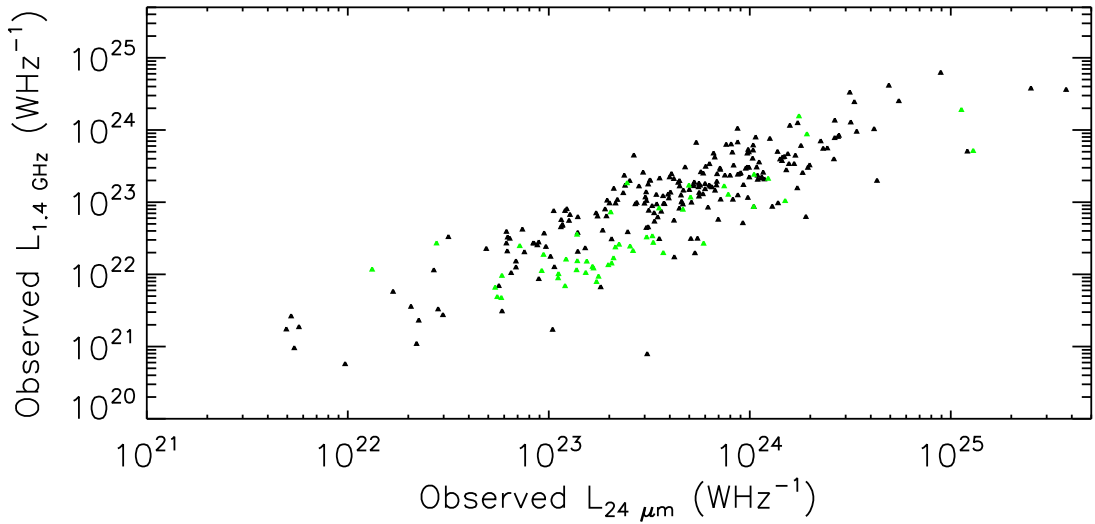} \caption{The luminosities 
at 1.4\,GHz and 24\,\mum\ for 224 24\,\mum\ \spitzer\ sources with redshifts and 
significant radio flux density from the HDF-N (black triangles) and 50 non-AGN 
24\,\mum\ sources with redshifts catalogued by Norris {\it et al.} (2006) (green 
triangles). The luminosities shown have not been k-corrected. All 1.4\,GHz 
luminosities have measured within an aperture of radius 1\farcs5 centred upon the 
\spitzer\ MIPS 24\,\mum\ source position. Note the few cases where HDF-N sources 
with apparently low radio luminosity with respect to MIR luminosity are large 
nearby spirals with an optical extent larger than the 3\arcsec.} 
\label{lumlum} %
\end{center} 
\end{figure*} 

\subsection{An evolution in the 
IR-radio correlation?}

It has been suggested, both theoretically \citep{bell03} and observationally 
\citep{boyle07}, that the IR-radio correlation may evolve at lower flux 
densities. One of the aims of this study is to investigate this correlation for 
very low flux density radio sources.

Derived from the HDF-N data alone the average value of q$_{24}$ shows a small 
trend to reduce as a function of lower values of S$_{\rm 24 \mu m}$ 
(Fig.\,\ref{q24_1} \& \ref{q24_1b}). This can be further demonstrated by splitting these data 
into two approximately equal samples, above and below S$_{\rm 24 \mu 
m}$=150\,$\mu$Jy. The mean and median values of q$_{24}$ for the 187 sources 
below 150\,$\mu$Jy are 0.45 and 0.36 respectively, compared with 0.56 and 0.54 
derived from the 190 sources with S$_{\rm 24 \mu m}>$150\,$\mu$Jy.  These HDF-N 
results along with results from samples of brighter galaxies 
\citep{appleton04,norris06} are broadly consistent with average q$_{24}$ reducing 
as a function of decreasing S$_{\rm 24 \mu m}$ (see Fig.\,\ref{q24_1} \& \ref{q24_1b}).
  
There are several contributing factors which will affect these results and those 
from previous studies. These factors include biases introduced by flux density 
thresholds of observations, the application or non-application of {\it k}-corrections 
and possible AGN contamination of sample galaxies.

Previous studies, such as \cite{appleton04}, have used samples of sources 
detected at both radio and infrared wavelengths. The implications of a radio 
detection limit on these results will be to exclude sources with high q$_{24}$ at 
low S$_{\rm 24\,\mu m}$. The line overlaid on Fig.\,\ref{q24_1} represents 
the upper limit imposed on the ATCA observations of the SWIRE survey by the radio 
detection threshold of these data \citep{norris06}. In a manner similar to that 
used by \cite{boyle07} no radio detection limit has been applied during analysis 
of these MERLIN+VLA data with the radio flux density being recorded for all MIR 
sources regardless of significance of their radio detection. This method will 
result in errors in the measured radio flux densities for very weak individual 
radio sources but will not directly bias against larger values of q$_{24}$. 
However when many sources are statistically averaged together the mean and median 
value of detected radio emission and hence the derived value of q$_{24}$ will be 
robust even for source considerably fainter than the radio detection threshold of 
these data.

No direct {\it k}-correction has been applied to these HDF-N flux density results 
because in many cases the spectral shapes and redshifts of objects are not known 
adequately enough to apply any correction with confidence. Additionally previous 
studies, such as by \cite{appleton04}, have shown that the application of 
k-corrections to individual sources below z=1 only introduces a small change to 
the average value of q$_{24}$ as a function of redshift. Of the sources with 
known redshifts in our sample 78 percent have redshifts less than 1, and thus will be 
minimally affected. Within Figs.\,\ref{q24_1} and \ref{q24_2} the values of 
q$_{24}$ for individual sources are plotted against their flux densities at each 
wavelength. These sources sample a wide range of luminosities and redshifts and 
hence the systematic effects resulting from not applying a {\it k}-correction 
will effect individual sources but will cause only minimal changes in any 
overall trend of the sample as a function of flux density. However, the 
effects of the application of a {\it k}-correction to these data do warrant further 
discussion.

Assuming that the 24\,\mum\ sources identified in this sample are predominately 
powered by star-formation, at an observed radio frequency of 1.4\,GHz the 
emission will be dominated by non-thermal synchrotron emission from supernovae 
and supernova remnants \citep{muxlow07}. As such the non-thermal radio signal 
will be boosted by a factor of (1+$z$)$^{0.7}$, assuming a synchrotron power law 
(S$\propto\nu^{-0.7}$). 
This {\it k}-correction 
will become incorrect for the highest redshift sources as the rest-frame 
radiation will be shifted to higher radio frequencies where the emission from 
other mechanisms such as thermal free-free emission will become increasingly 
important. For example a source at redshift of $\gtsim$2.5 the radio emission 
observed at 1.4\,GHz will have been emitted at a rest-frame frequency of 
$\gtsim$5\,GHz at which point an increasing proportion of the radio emission from 
a star-forming galaxy will arise from thermal processes resulting in a reduction 
in the radio spectral slope and hence {\it k}-correction that should be required.

Accurately {\it k}-correcting the \spitzer\ 24\,\mum\ flux densities requires detailed 
knowledge of the mid-IR spectral template of the individual sources. In 
particular, for sources at redshifts $>1$ significant spectral features from PAHs 
and silicates observed in local starburst galaxies \cite[e.g.][]{brandl06} are 
shifted into the \spitzer\ 24\,\mum\ band which can result in large additional 
{\it k}-correction factors. The inclusion of these line features can result in an 
increase of a factor of 2 or more in S$_{\rm 24\,\mu m}$ compared with commonly 
used starburst spectral templates \citep{fadda06,yan07}. This is equivalent to a 
change in q$_{24}$ of approximately 0.3. The general effect of the MIR 
{\it k}-correction, resulting from a spectral index of a starburst galaxy excluding 
line emission, will be to boost S$_{\rm 24\,\mu m}$ as a function of redshift by 
an amount that will depend upon spectral slope of the galaxy template used. For a 
luminous starburst this will result in a boosting with a MIR spectral slope 
$\alpha^{5}\!\!_{15}\sim 2$ \citep{yan07}. As such the effect of this 
k-correction applied to the 24\,\mum\ data will be larger than that applied to 
the 1.4\,GHz data, hence the overall effect will be to increase the value of 
q$_{24}$ as a function of redshift.

Uncorrected q$_{24}$ is plotted against redshift for the sources with known 
spectroscopic or photometric redshifts in Fig\,\ref{q24_3}. Applying {\it k}-corrections to these data points will 
result in a small increase in q$_{24}$ for the highest redshift sources plotted. 
Within the scatter of the small sample of sources plotted in Fig\,\ref{q24_3} no 
significant trend is observed as a function of redshift to z$\approx$3. This is 
consistent with previous studies \citep[e.g.][]{garrett02,appleton04}.

\subsubsection{Potential contamination from AGN}

In addition to instrumental and analysis effects outlined above it is possible 
that the flux densities of sources recorded in this study may include some 
proportion of contamination by embedded AGN \citep{richards07}. The effect of the 
inclusion of sources with significant AGN emission is demonstrated by the AGN 
sources from the SWIRE survey \citep{norris06} which are overlaid as stars in 
Figs.\,\ref{F20vsF24log}, \ref{q24_1} and \ref{q24_2}. In general, sources with a 
significant AGN contribution show an excess of radio emission compared to their 
infrared luminosity, hence will deviate from the IR-radio correlation. 
Consequently the result of any AGN contamination within the HDF-N \spitzer\ 
sample analysed here will be to reduce q$_{24}$. However, the flux densities of the HDF-N sample (see Fig.\,\ref{F20vsF24log}) 
show only a very small number of sources with excessive radio flux density which 
deviates significantly from the IR-radio correlation extrapolated from high flux 
density samples. This is to be expected since it is now generally accepted that 
below 100\,$\mu$Jy the vast majority of radio sources are predominantly powered by star-formation 
\citep{richards00,gruppioni03,seymour04,muxlow05}. Additionally this sample of 
sources is selected at 24\,\mum\ rather than at radio wavelengths and
hence will 
preferentially select star-forming sources rather than radio AGN. Thus, throughout 
this paper no direct distinction has been made between possible AGN and non-AGN 
star-forming galaxies in these faint \spitzer\ selected HDF-N radio sources. 
However, a small AGN contamination within the sources sampled cannot be 
eliminated. The consequences of any AGN contamination may, to some
degree, offer an explanation for the lower 
values of q$_{24}$ derived here \citep[see for example][]{daddi07}.

\subsection{Causes and implications of an evolution in the IR-radio correlation}

In the previous section a small trend in q$_{24}$ as a function of 24\,\mum\ flux 
density has been tentatively identified. This trend is small and only visible in 
statistically averaged deep radio observations. The much deeper radio surveys 
that will be made using the newly enhanced radio instruments such as {\it 
e-}MERLIN and the EVLA will constrain the faint end of the IR-radio correlation, 
confirming or refuting any possible deviations.

Whilst there are several observational and analysis effects which may contribute 
to the observed trend in the IR-radio correlation (outlined in the previous 
sections) there are also physical mechanisms which may cause an effect similar to 
that tentatively observed. Both the IR and radio emission from star-forming 
galaxies are indirectly linked with the rate of ongoing star-formation; 
simplistically the non-thermal radio emission originates in accelerated charged 
particles in supernovae and supernova remnants and the IR emission results from 
the reprocessing by dust of light from hot young stars. This inferred connection 
is generally considered to result in the IR-correlation. However since both the 
radio and IR emission are linked with star-formation via complex physical 
processes it is remarkable that the luminosity in these two bands correlates so 
accurately over such a large range of luminosities.  A recent study by 
\cite{bell03} has provided a detailed investigation of the relationship of the 
radio and total infrared emission of galaxies and their star-formation rates. In 
this work several competing physical factors which will affect the IR and radio 
emission from star-forming galaxies are considered which at low luminosities 
potentially cause a deviation in the IR-radio correlation similar to that which is 
very tentatively reported here.

If we assume that the non-thermal radio emission from star-forming 
galaxies is directly proportional to the star-formation rate (i.e. radio 
$\propto$ SFR), then the systematic reduction in q$_{24}$ at low 24\,\mum\ flux 
density implies that the 24\,\mum\ emission is no longer efficiently tracing the 
SFR of the galaxy \citep{bell03}. Observational evidence 
\citep[e.g.][]{wang96,adelberger00,hopkins01,buat02} has shown that galaxy 
luminosity and dust opacity to UV and H$\alpha$ emission are correlated, 
demonstrating that lower luminosity galaxies have substantially less dust 
absorption and reddening than high luminosity sources. These studies have also 
shown that the ratio of far-UV to IR light is significantly lower in less 
luminous sources, implying that whilst for luminous galaxies the IR provides a 
good tracer of star-formation at low luminosities much of the far-UV emission 
from star-formation is not reprocessed as IR emission \citep[see][for further 
discussion]{wang96,buat02,bell03}. The consequences of the reduced dust opacity 
in lower luminosity, presumable smaller, sources will be a reduction in the 
efficiency of reprocessing of starlight into IR emission and hence
will result in an observable 
decrease in the ratio of IR-to-radio emission from sources at low luminosities. 
Of course it should be reiterated that this makes the assumption that the radio 
emission is a `perfect' tracer of the star-formation which may equally not be 
true.

It has been suggested that the synchrotron radio emission from 
low-luminosity galaxies can also be suppressed, implying that it is not a 
true tracer of the SFR \citep{klein84,klein91, price92}. Whilst the connection 
between non-thermal radio emission and star-formation is complex, it has been 
discussed, both on the basis of models of the physical mechanisms and 
observational evidence, by many authors \citep[see for 
example][]{volk89,chi90,helou93}. A physical explanation for the suppression of 
the non-thermal radio emission in low-luminosity galaxies is that the 
accelerating electrons can escape these sources because they have less efficient 
cosmic-ray confinement than in larger, more luminous sources \citep{chi90}.

Radio surveys of IR sources prior to the very deep results presented here have 
not detected any significant deviations in the linearity of the radio-IR 
correlation \citep[e.g.][]{condon91,yun01,appleton04}. \cite{bell03} argue that 
this observed linearity at relatively low luminosities implies that both the IR 
and non-thermal radio emission do not adequately trace the star-formation in 
low luminosity sources. As such the observed linearity of the IR-radio 
correlation arises because the two independent physical mechanisms by which the 
emission in these bands originates both result in an underestimate of the SFR at 
low luminosities by a similar amount, thus preserving the observed linear nature 
of the correlation in a conspiratorial way \citep{bell03}. However, these two 
competing physical trends whilst similar in magnitude will diverge for very low 
luminosity sources resulting in a reduction in the IR-to-radio ratio. This effect 
can be seen in \citep[figures 3 and 8 from][]{bell03} and is tentatively hinted at 
in the observation presented here (Fig.\,\ref{q24_1}).

Not withstanding the factors, discussed in the previous section which may effect 
this tentative detection of a small deviation in the gradient of the IR-radio 
correlation, these observational results are broadly consistent with the models 
of \cite{bell03}. If this observed evolution of the IR-radio correlation at low 
flux densities is real it potentially has significant implications on the use of 
either IR or radio luminosities for the derivation of star formation rates of 
faint galaxies in future deep surveys.

\subsection{Future observational tests}

Future observations will be required to confirm this tentative deviation of the
IR-to-radio correlation at these extremely low flux densities.
These results may be tested further by {\it i)} deeper radio and
infrared observations which detect many faint individual
sources at high significance, {\it ii)} significantly larger area deep
surveys that allow robust statistical detections of faint sources to be
made, or {\it iii)} the separation and analysis of the various competing factors which contribute to
this effect. 

The stacked radio images of extremely faint IR sources presented in Fig.\,\ref{maps} 
provide a tantalising glimpse of the images which will be routinely available in 
the near future. The next-generation of radio 
interferometers, such as {\it e-}MERLIN \citep{garrington04} and the EVLA 
\citep{perley00}, will provide images many tens of times more sensitive than can 
currently be made.  Future deep radio observations using {\it e-}MERLIN and the 
EVLA with integration times comparable to those presented here will produce 
images with sub-microJy noise levels. Surveys using these instruments will easily be 
able to detect all of the IR sources so far found in the deepest \spitzer\ 
observations with very high confidence, and the high sensitivity of these instruments 
will allow much larger areas of microJy radio sky to be surveyed. Such
high-sensitivity and wide-field observations will constrain any low flux density 
deviations in the radio-IR correlation.

The radio observations presented here primarily trace the non-thermal synchrotron 
radio emission of galaxies with flux densities of a few tens of $\mu$Jy. At 
1.4\,GHz $\sim$90 percent of the emission of a typical $\sim$L$\ast$ spiral 
galaxy is non-thermal emission with the remainder coming from thermal 
bremsstrahlung processes \citep{condon92}. Whereas the radio synchrotron, as 
stated previously is indirectly related to star-formation, the bremsstrahlung emission 
is a more direct tracer of the ongoing star-formation since it arises from 
thermal radio emission from gas ionised by hot, young stars. At higher radio 
frequencies ($\gtsim$10\,GHz) thermal radio emission will begin to dominate over 
synchrotron emission \citep{condon92,price92} tending to flatten the radio 
spectrum of star-forming galaxies. Determining the relative fraction of thermal 
and non-thermal radio emission in different sources and at different frequencies 
is non-trivial requiring multi-frequency measurements of the radio flux densities 
in order to determine the slope of non-thermal emission.  Such observations will 
be more feasible using future extremely deep radio observations planned with the 
new generation of wide-bandwidth radio interferometers, such as {\it e-}MERLIN 
and the EVLA, which will allow sensitive multi-frequency radio observations of 
the distant Universe to be made within reasonable observation times. Using 
observations made with these enhanced centimetre instruments along with 
measurements made using longer wavelength radio interferometers like the Giant 
Metrewave Radio Telescope (GMRT) and LOw Frequency ARray (LOFAR) it will be 
possible to constrain the radio spectral energy distribution of many star-forming 
galaxies hence allowing the thermal and non-thermal radio emission to be 
separated. By disentangling the thermal and non-thermal radio luminosities of 
many distant star-forming galaxies it will be possible to verify the expected 
suppression of IR emission from low luminosity sources relative to the thermal 
radio emission which is a more direct, extinction free, tracer of the star-formation.

\section{Conclusions}

Using one of the deepest high-resolution 1.4\,GHz observations made to date, 
in conjunction with deep 24\,\mum\ \spitzer\ source catalogues from GOODS, we have 
investigated the microJy radio counterparts of faint MIR sources. These 
observations confirm that the microJy radio 
source population follow the MIR-radio correlation and extend this correlation by several orders of magnitude to 
very low flux densities and luminosities, and out to moderate redshifts. This extension 
of the MIR-radio correlation confirms that the majority of these
extremely faint radio and 
24\,\mum\ sources are predominantly powered by star-formation with little AGN 
contamination.

Statistically stacking the radio emission from many tens of faint
24\,\mum\ sources has been used to characterise the size and nature of the radio 
emission from very faint IR galaxies well below the nominal radio sensitivity of 
these data. Using these methods the MIR-radio correlation has been further 
extended and a tentative deviation in this correlation at very low 
24\,\mum\ flux densities has been identified.

\subsection*{Acknowledgments}

The authors wish to thank Ray Norris, Brian Boyle, Phil Appleton and Nick Seymour for 
extremely useful discussions. RJB acknowledges financial support by the European 
Commission's I3 Programme ``RADIONET'' under contract No. 505818.  This work is based 
on observations made with MERLIN, a National Facility operated by the University of 
Manchester at Jodrell Bank Observatory on behalf of STFC, and the VLA of the National 
Radio Astronomy Observatory is a facility of the National Science Foundation operated 
under cooperative agreement by Associated Universities, Inc. This work is based in 
part on archival data obtained with the Spitzer Space Telescope, which is operated by 
the Jet Propulsion Laboratory, California Institute of Technology under a contract 
with NASA. Support for this work was provided by an award issued by JPL/Caltech.

\bibliographystyle{mnras}

\end{document}